# Time Evolution of Disease Spread on Networks with Degree Heterogeneity

Pierre-André Noël,[1, 2]
Bahman Davoudi,[1]
Louis J. Dubé,[2, 3]
Robert C. Brunham[1]
Babak Pourbohloul[1, 2, 4, *]

[1]University of British Columbia Centre for Disease Control,
Vancouver, British Columbia, Canada V5Z 4R4
[2]Département de Physique, de Génie Physique et d'Optique,
Université Laval, Québec, Québec, Canada G1K 7P4
[3]Laboratoire de Chimie Physique-Matière et Rayonnement,
Université Pierre et Marie Curie, 75231 Paris Cedex 05, France
[4]Department of Health Care & Epidemiology,
University of British Columbia, Vancouver,
British Columbia, Canada V6T 1Z4

(Date: Oct 17, 2007)

[*]Corresponding Author:

Babak Pourbohoul, PhD
Division of Mathematical Modeling
UBC Centre for Disease Control
655 West 12[th] Avenue, Vancouver, BC
Canada   V5Z 4R4

**Abstract**

Two crucial elements facilitate the understanding and control of communicable disease spread within a social setting. These components are, the underlying contact structure among individuals that determines the pattern of disease transmission; and the evolution of this pattern over time. Mathematical models of infectious diseases, which are in principle analytically tractable, use two general approaches to incorporate these elements. The first approach, generally known as compartmental modeling, addresses the time evolution of disease spread at the expense of simplifying the pattern of transmission. On the other hand, the second approach uses network theory to incorporate detailed information pertaining to the underlying contact structure among individuals. However, while providing accurate estimates on the final size of outbreaks/epidemics, this approach, in its current formalism, disregards the progression of time during outbreaks. So far, the only alternative that enables the integration of both aspects of disease spread simultaneously has been to abandon the analytical approach and rely on computer simulations. Powerful modern computers can perform an enormous number of simulations at an incredibly rapid pace; however, the complex structure of "realistic" contact networks, along with the stochastic nature of disease spread, pose serious challenges to the computational techniques used to produce robust, real time analysis of disease spread in large populations. An analytical alternative to this approach is lacking. We offer a new analytical framework based on percolation theory, which incorporates both the complexity of contact network structure and the time progression of disease spread. Furthermore, we demonstrate that this framework is equally effective on finite- and "infinite"-size networks. Application of this formalism is not limited to disease spread; it can be equally applied to similar percolation phenomena on networks in other areas in science and technology.

The spread of communicable diseases is a dynamical process and as such, understanding and controlling infectious disease outbreaks and epidemics is pertinent to the temporal evolution of disease propagation. Historically, this aspect of disease transmission has been studied with the use of "coarse-grained" dynamical representation of populations, known as compartmental models.[1, 2] In these models, a population is divided into a number of epidemiological "states" (or classes) and the time evolution of each is described by a differential equation. Figure 1a shows a schematic diagram of a simple Susceptible-Exposed-Infected-Removed (SEIR) model, in which every individual can be in the susceptible, exposed, infected or removed class at any given time. Although this approach, and its more complex variants, has been instrumental in understanding several features of infectious diseases over the past 3 decades, it comes with a major simplification. The simplifying assumption states that the population is "well mixed", i.e., every individual has an equal opportunity to infect others. This assumption may be valid in the broader context of population biology. Human populations, however, tend to contact each other in a heterogeneous manner based on their age, profession, socio-economic status or behavior, and thus, the well-mixed approximation cannot portray an accurate image of disease spread among humans specifically in finite-size populations.[3] Recent advances in network- and percolation- theories, have paved the way for physicists to bring a new perspective to understanding disease spread. Over the past decade, seminal work by Watts and Strogatz on small-world networks[4], Barabasi *et al.* on scale-free networks[5] and Dorogovtsev, Mendes[6], Pastor-Satorras and Vespignani[7] among others on the dynamics of networks has shed light on a number of intriguing aspects of epidemiological processes. In particular, groundbreaking work by Newman *et al.*[8, 9, 10] has provided a strong foundation for the formulation of epidemiological problems using tools developed by physicists.

Contrary to compartmental approaches, a network representation of a system takes into account that each individual does not have the same probability of interacting with every other individual; in fact, one interacts only with their "topological" neighbours (figure 1b). Taking this into account, we map a system of $N$ individuals to a network in which each individual is represented by a node (vertex) and the connection between each pair of individuals is represented by an edge (link). We call part of the link that is connected to a vertex, a stub. Each vertex has a degree $k$ (number of neighbours) and the set $\{k_i\}$ (called degree sequence) partially defines the network.

In many practical situations, the degree sequence is the only available information and our best guess for the network structure is one in which the $k_i$ stubs of vertex $i$ are randomly connected to stubs of the other vertices (with no self-loops), while respecting the degree sequence. It is common to explicitly forbid two vertices to share more than one link (simple graph). For sparse graphs - in which the number of links scales linearly with the number of nodes - the probability for such an event decreases as $1/N$ and can be neglected for large networks.[8]

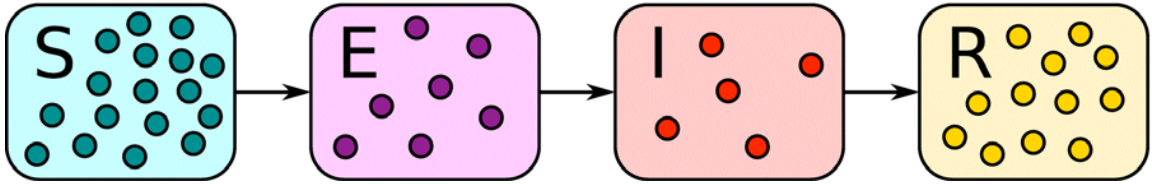

**Figure 1a: A simple SEIR compartmental model.** At any moment, individuals (circles) can be part of the population within any of the following compartments: *S*usceptibles (not infected but could potentially become so upon interaction with infectious), *E*xposed (carries the disease but cannot transmit it), *I*nfectious (can transmit the disease) or *R*emoved (cannot transmit the disease nor become newly infected). Individuals within a compartment are indistinguishable and thus flow among compartments depends solely on the total population of compartments. More complex variations exist where, for example, these compartments are divided in sub-compartments that define more specific groups of individuals, which allows more detailed dynamics.

**Figure 1b: A network model.** Each individual is represented by a vertex in a network. Neighbouring vertices are vertices linked by an edge. A stub is the half of an edge that is connected to a given vertex. The disease can only be transmitted to neighbours of an infectious vertex (but not necessarily to all of them). Neighbours do not necessarily need to be geographical neighbors. For instance, if an individual does not come into contact with his/her household neighbours, but does come into contact with someone in a shopping mall (not necessarily in the household neighborhood) he/she may infect these "topological" neighbours. In the limit of an infinite population, classic network models can give an analytical probability distribution for outbreak size; or the probability and size of a large-scale epidemic. However, unless massive numerical simulations are performed, they do not bring much information about the temporal evolution of the disease other than information on its final state.

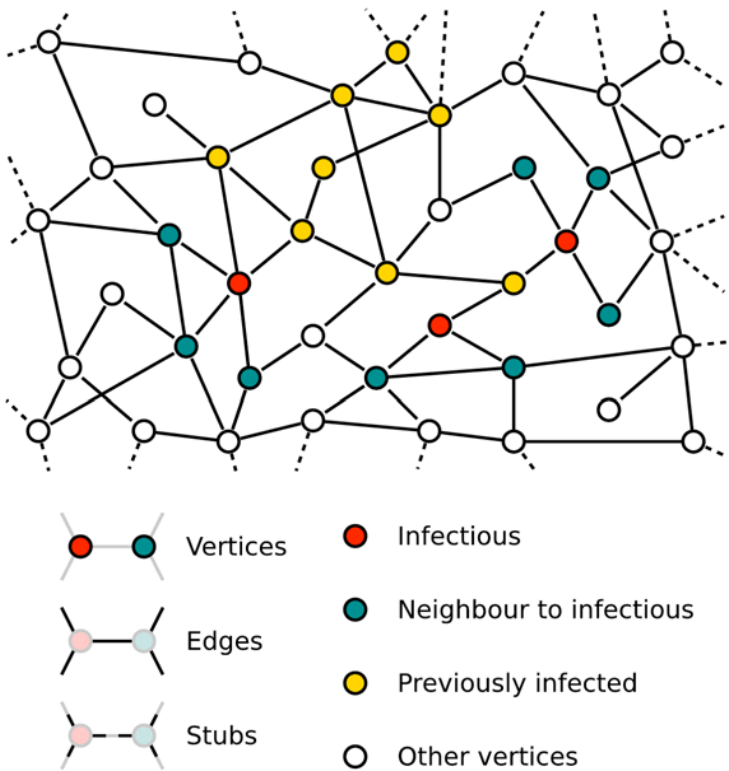

We are primarily interested in diseases where infected individuals are eventually removed from the dynamics of the system (i.e., infection is followed by naturally-acquired immunity or death); thus, the same person cannot be infected more than once. At any given time, we call an individual "susceptible" if (s)he has never been exposed to the disease; "exposed" if (s)he has acquired the infection but not currently able to pass on the disease to another person; "infectious" if (s)he is currently able to transmit the infection to others; removed if (s)he became immune or succumbed to death after acquiring the infection; and finally, "infected" if (s)he has been exposed to the infectious agent at least once in the past, regardless of their current state (e.g. exposed, infectious, removed). Disease transmission can only occur between immediately linked individuals.

Network analysis using the generating function formalism, developed by Newman *et al.*, is a powerful tool when analyzing the spread of disease within networks.[9] Without directly addressing the question of "when the transmission occurs?", it provides reliable results on the final size of an outbreak/epidemic by addressing the question of "whether transmission occurred?". To address the first question, i.e., the time evolution of the system, we design a mapping between time and disease generation and show how this can be used in the generating function formalism. Using mean-field approximations, we then obtain corrections to take into account the impact of the finite size of a network on the epidemic process.

In recent years, several researchers have recognized the importance of incorporating the notion of time into the generating function formalism that describes percolation dynamics on networks. In broaching this issue, many notable advances have been made. Recently, one group addressed the probability distribution of outbreak sizes as a function of time for infinite-size networks[11], while another group addressed the finite-size effect by deriving estimates on the mean size of a large-scale epidemic, rather than the probability distribution. [12, 13] As we will show in the subsequent sections, despite these advances, one is required to develop a truly integrated analytical framework that encompasses both the time progression of disease and the network finite-size effect.

Although we focus specifically on disease spread as the dynamical phenomenon in this paper, the methodology is quite general and can be applied to other processes that manifest similar dynamical properties.

## **Basic Generating Function Formalism**

With knowledge of the degree sequence of the physical or social network of interest, we can obtain the set of probabilities $\{p_k\}$ that a random vertex has degree $k$ (degree distribution). Following Newman et *al.*[8] and Newman[9], we define the probability generating function (*pgf*) for a random vertex $G_0(x) = \sum_{k=0}^{\infty} p_k x^k$, respecting $G_0(1) = \sum_k p_k x^k = 1$ when $p_k$ is properly normalized. The average degree, $z_1$, can be easily obtained from $z_1 = \langle k \rangle = \sum_{k=0}^{\infty} k p_k = G_0'(1)$, where the prime denotes the derivative with respect to $x$. They also showed the probability, $q_k$, that $k$ vertices could be reached from a vertex we arrived at by following a random edge (excluding this edge from the count), can be derived as $G_1(x) = \sum_{k=0}^{\infty} q_k x^k = \frac{\sum_k (k+1) p_{k+1} x^k}{\sum_k (k+1) p_{k+1}} = \frac{1}{z_1} G_0'(x)$.

While $G_0(x)$ and $G_1(x)$ contain information about the structure of the physical network linking individuals within the epidemiological system, they do not hold any information about the risk of disease transmission between two neighbouring vertices. However, with the additional knowledge of the transmissibility, $T$ – the probability that an infectious node will eventually infect one of their neighbours – one can determine the probability of infecting $m$ out of $k$ neighbours, $\binom{k}{m} T_0^m (1-T_0)^{k-m}$. We can thus define the *pgf* for the number of infections directly caused by the initial infectious case ("patient zero") as

$$G_0(x;T_0) = \sum_{m=0}^{\infty} \sum_{k=m}^{\infty} p_k \binom{k}{m} T_0^m (1-T_0)^{k-m} x^m$$
$$= G_0(1 + (x-1)T_0).$$

(1)

One can continue in the same vein and obtain informative results about the final state of the population, after the outbreak/epidemic has ended[9, 3]; however, this approach in itself does not yield any new information about the duration of the epidemic, its speed of propagation or other time-related quantities.

## **Temporal Interpretation of Infection Transmission**

To circumvent the temporal limitations of the generating function formalism presented in the previous section, we adopt an approach focusing on generations of infection. We define a generation of infection as the mean time between an individual becoming infected and passing on their infection to others. We, thus, define *generation 0* as the initial patient (index, or patient zero) and individuals of generation $g$ as those who acquired the disease from a member of generation $g-1$. Since disease transmission occurs solely between neighbouring individuals, generation $1$ is thus composed of immediate neighbours of patient zero; in the same manner, generation $2$ is part of the neighbourhood of generation $1$, and so on. However, not every neighbour of an infectious individual will become infected; thus, we define $T_g$ as the probability that transmission occurs between any individuals of generation $g$ and each of their susceptible neighbours. The special case where this transmissibility is the same for every generation, $g$, (i.e., $T_g = T$) is called the *stationary* case and corresponds to a situation where the parameters governing the dynamics of the disease do not vary in time.

Non-stationarity of a system can come from intrinsic properties (e.g. seasonality) or exposure to external interventions (e.g. vaccination campaign, treatment, modification of behaviours, etc.). There is clearly a causality link among generations and one would expect individuals of higher generations to, on average, become infected later than those in generations closer to patient zero. This is particularly clear for the special "algorithmic" case where transmission occurs between succeeding generations at constant time intervals of $\tau$. In this scenario, time $0$ is defined as the time of infection for patient zero and generation $g$ becomes infected at time $g\tau$. However, no known real-world disease follows this idealistic behaviour and a more general formalism must be established to achieve this mapping. In this section, we use an approach closely related to linear combination of impulse responses to achieve this goal.

## A. Infection Rate

The time evolution of disease spread in an epidemiological system has been extensively studied using either compartmental[1, 2, 14, 15, 16] or agent-based[17, 18, 19] models. In both cases, it is important to know the disease transmission rate from individual $i$ to $j$. In an SEIR compartmental model, similar individuals are regrouped in different classes (compartments), namely susceptible, exposed, infectious and removed; the populations within each compartment at time $t$ are denoted by $S(t)$, $E(t)$, $I(t)$ and $R(t)$, respectively. Their time dependence can be obtained from the differential equations system

$$\dot{S} = -\lambda(t)SI; \quad \dot{E} = \lambda(t)SI - \alpha(t)E; \quad \dot{I} = \alpha(t)E - \mu(t)I; \quad \dot{R} = \mu(t)I \qquad (2)$$

where dot denotes derivative with respect to time; $\alpha(t)$ and $\mu(t)$ define the incubation and infectious periods, respectively; and $\lambda(t)$ identifies the probability rate of an infectious individual infecting a susceptible individual.[2] With appropriate normalization, one can change the perspective of the problem slightly and define $S_i$, $E_i$, $I_i$ and $R_i$ as the probabilities for each individual, $i$, to be in the corresponding compartment. With the knowledge that the $i$-th individual has been exposed to the disease at time $t^*$, the above set of equations become

$$\frac{dE_i}{dt} + \alpha(t)E_i = \delta(t - t^*)$$

$$\frac{dI_i}{dt} + \mu(t)I_i = \alpha_i(t)E_i \qquad (3)$$

with $\delta(t)$ being the Dirac delta function. The probability that this individual becomes infectious at time $t$ can be found by solving the abovementioned equations, subsequently yielding

$$I_i(t|t^*) = \Theta(t - t^*)e^{-\int_{t^*}^{t}\mu(t'')dt''}\int_{t^*}^{t}\alpha(t')e^{\int_{t^*}^{t'}(\mu(t'')-\alpha(t''))dt''}dt' \qquad (4)$$

where $\Theta(t)$ is the Heaviside function. The disease transmission rate is thus given by $r_{ij}(t|t^*) = \lambda_{ij}(t)I_i(t|t^*)$. Such a quantity can either be obtained from a mathematical model, as in the previous example, or by direct observation of a biological system. By definition, $r_{ij}(t|t^*) = 0$ for $t < t^*$ as node $j$ cannot be infected from node $i$, if $i$ is not yet infected (causality). Generally, the infection rate of a disease obeying an SEIR like dynamic is typically close to zero in the immediate time interval after $t^*$, then increases to a maximum value and finally decreases to zero.

## B. Transmissibility

We previously defined transmissibility, $T_g$, as the probability for any infectious individual in generation $g$ to *eventually* infect one of their susceptible neighbours. In this section, we extend this concept and define the *disease transmission probability*, $T_{ij}(t|t^*)$, as the probability that an individual known to be infected since time $t^*$, has infected one of their neighbors before time $t$. Recalling that an individual can be infected only once, one can obtain $T_{ij}(t|t^*) = 1 - \exp\left(-\int_{t^*}^{t} r_{ij}(t'|t^*)dt'\right)$. The transmissibility defined in the previous section is the ultimate disease transmission probability ($t \rightarrow \infty$) and can be written as $T_{t^*} = T(\infty|t^*)$.

It has been shown that when the $r(t|t^*)$ is not identical for every pair of individuals, but is rather an independent and identically-distributed (iid) random variable, then $T(t|t^*)$ obeys the same statistics.[9] This means the *a priori* transmissibility for a randomly selected individual is simply the average $T(t|t^*)$ over its distribution along all edges in the network. Thus, the fate of transmission along an edge does not depend on the detailed behaviour of $r(t|t^*)$ (e.g., incubation period), but rather it depends on the total area under the curve $r(t|t^*)$. Ultimately, it is the degree of infectivity and the duration of infectious period that are of utmost importance. To investigate further, we define the *infection time distribution*, $\phi(t|t^*) = \frac{1}{T_{t^*}}\frac{d}{dt}T(t|t^*)$, such that the probability of infection during the interval [t, t+dt] is given by $\phi(t|t^*)dt$. Therefore, the disease transmission probability can be written as

$$T(t|t^*) = T_{t^*}\int_{t^*}^{t}\phi(t'|t^*)dt'.$$ (5)

We call $r(t|t^*)$ *stationary* if $r(t|t^*) = r(t - t^*)$ for all $t^*$; this occurs in an SEIR model when the parameters $\lambda$, $\alpha$ and $\mu$ in equation (2) are time independent. From the definition of transmissibility we have $T(t|t^*) = T(t - t^*)$, which implies that the disease transmission probability is stationary and depends solely on the elapsed time since $t^*$. Likewise, the latter can be said for $\phi(t|t^*) = \phi(t - t^*)$ and $T(\infty|t^*) = T$. This is in agreement with the definition of stationarity presented in the previous section.

In general, members of the same generation do not all become infected at the same time. However, knowing that the initial patient is infected at time 0, we define $\phi_0(t)$, the infection time distribution for generation 1, as $\phi_0(t) = \phi(t|0)$. For higher generations the recurrence $\phi_g(t) = \int_0^\infty \phi(t|t')\phi_{g-1}(t')dt'$ can be used to obtain the infection time distribution for the transmission from an individual in generation $g$. The disease transmission probability for an individual in generation $g$ is thus given by

$$T_g(t) = \int_0^\infty T(t|t')\phi_{g-1}(t')dt'. \tag{6}$$

This relationship defines the temporal variation of transmissibility in an exact manner, and allows us to find $T_g = T_g(\infty)$. In the stationary case, equation 6 is in fact a convolution function and additional information about stationary and quasi-stationary cases are presented in the Supplementary Online Material.

## Time-dependent dynamics on an Infinite-size Network

We now use the results in the previous sections and expand the formalism to introduce a new *pgf* for an arbitrary generation $g$

$$G_g(x;T_g) = \begin{cases} G_0(1 + (x-1)T_0) & for\ g = 0 \\ G_1(1 + (x-1)T_g) & for\ g \geq 0 \end{cases}. \tag{7}$$

This will generate the probability distribution for the number of vertices that acquire infection directly from a single vertex of generation $g$ (for all equations within this section $T_g(t)$ can directly substitute $T_g$). In the stationary case ($T_g = T$ for all $g$) of Newman's formalism, the terms corresponding to $g = 0$ and $g = 1$ in the above generating function are identical to their counterparts, $G_0(x; T)$ and $G_1(x; T)$, respectively. From the properties of the *pgf*'s, the expected number of secondary infections caused directly by an infected individual in generation $g$ is given by

$$\langle m_g \rangle = \frac{dG_g(x;T_g)}{dx}\bigg|_{x=1} = \begin{cases} T_0 G_0'(1) & for\ g\ =\ 0 \\ T_g G_1'(1) & for\ g\ \geq\ 0 \end{cases}. \tag{8}$$

In the stationary case, $\langle m_g \rangle$ is identical for every generation except the first one and corresponds to the *basic reproductive number* $R_0 = TG_1'(1) = T\dfrac{z_2}{z_1}$, where $z_2 = G_0''(1)$ is the expected number of second neighbours for a random node. $R_0$ corresponds to the expected ratio of infected individuals in two successive generations (excluding $g = 0$); furthermore, $R_0 < 1$ implies that the expected number of infected individuals decreases in consecutive generation, leading to the extinction of the disease. Conversely, $R_0 > 1$ implies that the expected number of infected individuals increases and can potentially lead to an epidemic, which is a *giant component* of occupied edges.[3] It is worth noting that $R_0 > 1$ does not guarantee the occurrence of an epidemic; a finite number of realizations below the expectation can still lead to the extinction of the disease. Similar results are observed in non-stationary cases with the exception that $R_0$ is allowed to vary with generations.

The *pgf*'s $G_g$ $(x; T_g)$ defined above hold when the number of infected individuals in the current and previous generations are small compared to the size, $N$, of the network; in such a case, the probability of infecting an individual that is already infected is proportional to $N^{-1}$. This condition is fulfilled either when there is no giant component or when we limit ourselves to the first few generations. We assume that a person cannot be infected twice and the propagation of the disease follows a tree-like structure (without a closed loop). This allows us to define $H_g^h(x; T_g,...,T_{h-1})$, with $g < h$, the generating functions for the number of infected individuals that originates from a given individual of generation $g$ (including that individual) up to generation $h$. Clearly, $H_{h-1}^h(x; T_{h-1}) = xG_{h-1}(x; T_{h-1})$ holds for all adjacent generations. Using the properties of generating functions, the number of infections that emanate from $m$ different nodes of generation $g + 1$ (including these nodes), up to generation $h$, is generated by $[H_{g+1}^h(x; T_{g+1},...,T_{h-1})]^m$. We thus have the (backward) recurrence equation,

$$
\begin{aligned}
H_g^h(x; T_g,...,T_h) &= \sum_{m=0}^{\infty} \sum_{k=m}^{\infty} p_k \binom{k}{m} T_g^m (1-T_g)^{k-m} \left[ H_{g+1}^h(x; T_{g+1},...,T_{h-1}) \right]^m, \\
&= G_g(H_{g+1}^h(x; T_{g+1},...,T_{h-1}); T_g)
\end{aligned}
\tag{9}
$$

from which we can obtain the generating function for the total size of the epidemic up to generation $g$: $H_0^g(x; T_0,....T_{g-1}) = xG_0(xG_1(xG_2(...G_{g-1}(x; T_{g-1})...); T_1); T_0)$. Moreover, in the stationary case ($T_g = T$), we have

$$
\begin{cases}
H_0^\infty(x; T) = xG_0(xG_1(xG_1(...); T); T) \\
\qquad\quad = xG_0(H_1^\infty(x; T); T) \\
H_1^\infty(x; T) = xG_1(H_1^\infty(x; T); T)
\end{cases}
\tag{10}
$$

which is in agreement with previous results. Once $H_0^g(x; T_0,...,T_{g-1})$ is known, the properties of the generating functions can be utilized to derive the expected size of an outbreak after $g$ generations: $\langle s_g \rangle = \frac{d}{dx} H_0^g(x; T_0,...,T_{g-1}) \Big|_{x=0}$.

In the presence of a giant component, two qualitatively different outcomes (small outbreaks or large-scale epidemics) are possible. In such a case, the average value, $\langle s_g \rangle$, lies somewhere between these outcomes and does not confer much information about the risk of disease spread in each situation, separately. More generally, we define $P_{s,g}$ as the probability of observing the total number of $s$ infected individuals after $g$ generations.

This can be obtained from the derivatives of $H_0^g(x;T_0,...,T_{g-1})$ evaluated at $x = 0$; more specifically, $P_{s,g} = \dfrac{1}{s!}\dfrac{d^s H_0}{dx^s}\bigg|_{x=0} = \dfrac{1}{2\pi i}\oint\dfrac{H_0(z;t)}{z^{s+1}}dz$.

As for a finite time, $t$, there is always a $g$ in which $T_{g-1}(t)$ is negligibly small; one can use $H_0^g\big(x;T_0(t),...,T_{g-1}(t)\big)$ to obtain the probability distribution for the size of the outbreak/epidemic at that time.

Although a systematic derivation of generating functions required the introduction of equation (9), its application may be limited due to the backward nature of its recurrence relation. This means the results for generation **g-1** cannot be used for generation **g** and each $H_0^g(x;T_0,...,T_{g-1})$ must be obtained independently. To overcome this limitation, the next section introduces a formalism based on a forward recurrence relation. This formalism also proves advantageous when including networks with finite-size effects.

**Phase-space Representation**

It is convenient to define $\psi_{s,m}^g$ as the probability of having *s* infected individuals by the end of the **g**-th generation, of which **m** were infected during the **g**-th generation. This probability is generated by $\Psi_0^g(x,y) = \sum_{s,m}\psi_{s,m}^g x^s y^m$, from which we can obtain the (forward) recurrence relation:

$$\Psi_0^g(x,y) = \sum_{s',m'}\psi_{s',m'}^{g-1}x^{s'}\Big[G_{g-1}\big(1-(xy-1)T\big)\Big]^{m'} \tag{11}$$

with the initial condition $\Psi_0^0(x,y) = xy$ (or $\psi_{s,m}^0 = \delta_{s,1}\delta_{m,1}$, where $\delta_{i,j}$ is the Kronecker delta). This recurrence can also be expressed as $\Psi_0^g(x,y) = \Psi_0^{g-1}(x,G_{g-1}(xy;T_{g-1}))$. From $\Psi_0^g(x,y)$ we can obtain the generating function, $G_0^g(x;T_0,...,T_{g-1})$, for the number of infections among generation **g** in addition to the generating function for the total outbreak size

$$\begin{aligned}G_o^g(x;T_0,...,T_{g-1}) &= \Psi_0^g(1,x)\\H_o^g(x;T_0,...,T_{g-1}) &= \Psi_0^g(x,1)\end{aligned}. \tag{12}$$

These are in fact simple projections of the matrix $\psi_{s,m}^g$ on its respective *s* and **m** dimensions. Each element of this (triangular) matrix can be seen as a possible "state of infection" where the *s* and **m** dimensions provide information about the "position" (number of infected) and "momentum" (new infections) in the infection space, respectively. Figure 2 demonstrates the phase space and the projection on the *s* axis for a power law distribution for *N* = 1,000.

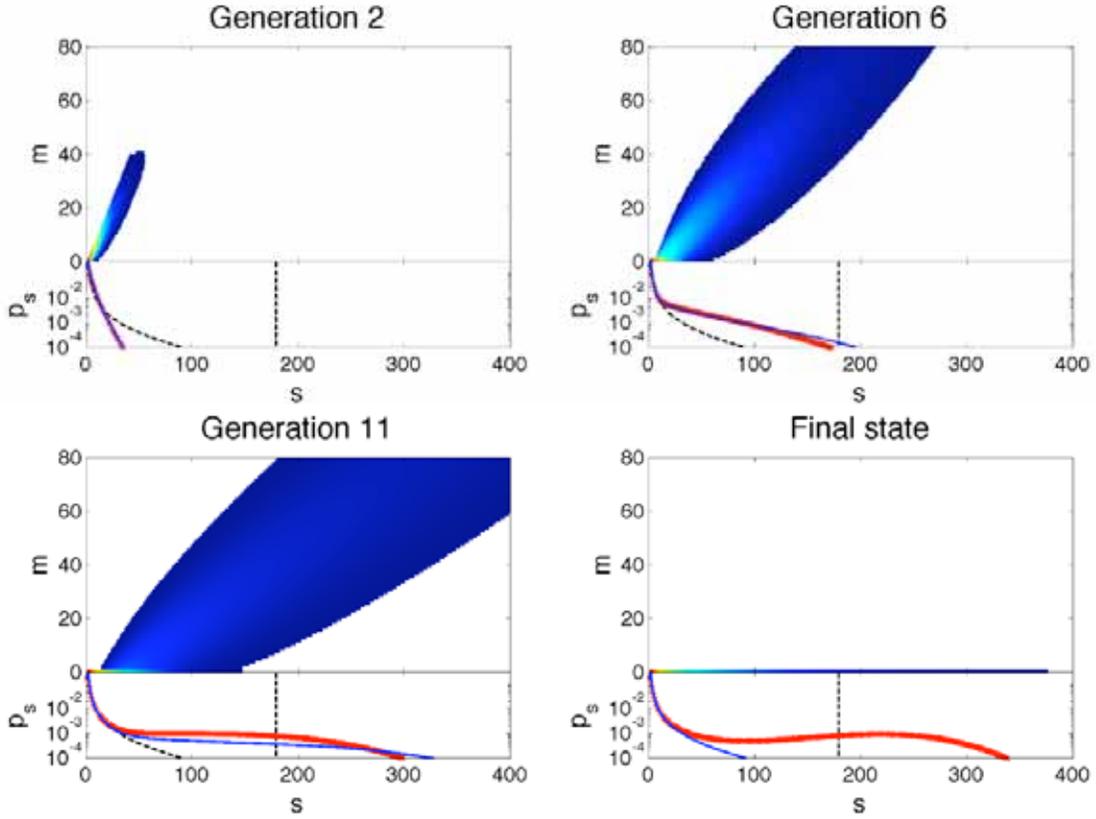

**Figure 2: Phase-space representation for the infinite-size network algorithm.** The degree distribution of the $N = 1,000$ individuals follows a power-law with $\tau = 2$ and $\kappa = 5$ (see Table 1) and the probability of transmission along an edge is $T = 0.8$. The phase-space representation for generations 2, 6 and 11 as well as for the final state are shown together with the corresponding projection on the $s$ axis (solid blue lines). The corresponding numerical results (red crosses), theoretical "infinite-size" outbreak distribution (dashed black curves) and theoretical "infinite-size" epidemic size (vertical black dashed lines) are also displayed. Numerical results are obtained by creating an ensemble of 1,000 equivalent graphs, each of which was used to run 10,000 simulations, performing 10 million epidemic simulations on all networks in total. The figures clearly demonstrate that apart from the small-scale outbreaks, the results from the infinite-size formalism may not correctly predict the outbreak/epidemic size distribution for finite-size networks. The remedy to this shortcoming is offered in the next section.

## Finite-size Effects

So long as one is only interested in the initial stage of an outbreak, the "infinite-size" assumption has negligible effects on the dynamics of disease spread. However, large deviations become evident when a sizeable fraction of the network has been affected. While the size of small outbreaks is mostly governed by stochastic fluctuations, the size of the giant component (when one exist) is limited by two principal causes: the evolution of the degree distribution of susceptible individuals; and the failure of transmission due to the impossibility of re-infection.

As equation (11) is *exact* in the infinite limit, one can search for a similar form where the finite-size effects are introduced as a dependency in $s'$ and $m'$ of the degree distribution and/or of its parameters. The following describes in more details these effects and how they are introduced into the model.

## A. Evolution of the Degree Distribution of Susceptibles

As the disease progresses across the network, susceptible individuals with a higher degree of connectivity are more likely to acquire the disease than those with fewer connections. If one focuses only on the degree distribution of susceptible cases, the distribution will vary over time; the portion representing high-degree susceptibles will decrease and the segment representing low-degree susceptibles will increase, to comply with normalization requirements. This variability over time has a direct effect on the ratio $\frac{z_2}{z_1}$ and can lower the reproduction number, $R$, below the threshold value of 1; consequently, leading to extinction of the disease albeit there is minimal depletion in the pool of susceptible individuals. This effect is particularly important for degree distributions in which some individuals have a degree much higher than the mean degree distribution (e.g., power-law distribution). The removal of these individuals will significantly impact the connectivity of the network, thereby diminishing the likelihood of establishing a giant component.

To take this effect into account we define $G_0^S(x;s)$, the generating function for the degree distribution of the remaining susceptibles for the current size of the outbreak/epidemics, $s$. By solving a differential equation in the mean-field approximation, one can show that this degree distribution would be (see Supplementary Online Material):

$$p_k^S(s) = p_k \frac{N-1}{N-s} (\theta(s))^k, \qquad (13)$$

where $\theta(s)$ satisfies the condition $\sum_k p_k \left(\theta(s)\right)^k = \frac{N-s}{N-1}$. Table I shows $\theta(s)$ for some typical distributions.

| Degree Distribution | Probability function | $\theta(s)$ |
|---|---|---|
| Poisson | $p_k = \dfrac{e^{-z}z^k}{k!}$ | $\theta(s) = \dfrac{1}{z}\left[ z + \ln\left( \dfrac{N-s}{N-1} \right) \right]$ |
| Binomial | $p_k = \dbinom{N}{k} p^k (1-p)^{N-k}$ | $\theta(s) = \dfrac{1}{p}\left[ \left( \dfrac{N-s}{N-1} \right)^{1/N} + p - 1 \right]$ |
| Exponential | $p_k = (1 - e^{-1/\kappa})e^{-k/\kappa}$ | $\theta(s) = \dfrac{N-1-(s-1)e^{1/\kappa}}{N-s}$ |
| Power Law | $p_k = \dfrac{k^{-\tau}e^{-k/\kappa}}{Li_\tau\left(e^{-1/\kappa}\right)}$ , for $k \geq 1$ | $Li_\tau\left(e^{-1/\kappa}\theta(s)\right) = \dfrac{N-s}{N-1}Li_\tau\left(e^{-1/\kappa}\right)$ |

Table I: Expression for $\theta(s)$ for some common degree distributions.

It is worth noting that for the Poisson distribution, this process preserves the shape of the distribution while shifting its average to a lower value $z_1(s) = z + \ln\left(\dfrac{N-s}{N-1}\right)$.

When an analytical closed form satisfying equation (13) cannot be found for $\boldsymbol{\theta(s)}$, the quantity $p_k^S(s)$ can be derived numerically for each pair of $\boldsymbol{k}$ and $\boldsymbol{s}$. Once $G_0^S(x;s)$ is known, one can show that the degree distribution of the susceptibles in the previous generation is given by $G_0^S(x;s-m)$. Subtraction and proper normalization thus yield the degree distribution, $G_0^I(x;s,m)$, of those that became infectious in the last generation

$$p_k^I(s,m) = \frac{(N-s+m)p_k^S(s-m) - (N-s)p_k^S(s)}{m}. \tag{14}$$

The excess degree of the currently infectious individuals is therefore generated by

$$G_1^I(x;s,m) = \frac{G_0^I(x;s,m) - G_0^I(0;s,m)}{x\left(1 - G_0^I(0;s,m)\right)}. \tag{15}$$

The distribution can be used in equation (11) as an approximation of $\boldsymbol{G_g(x)}$ when finite-size effects cannot be neglected.

Figure 3 shows the variation of degree distributions of susceptible and infected individuals for two of stylized distributions.

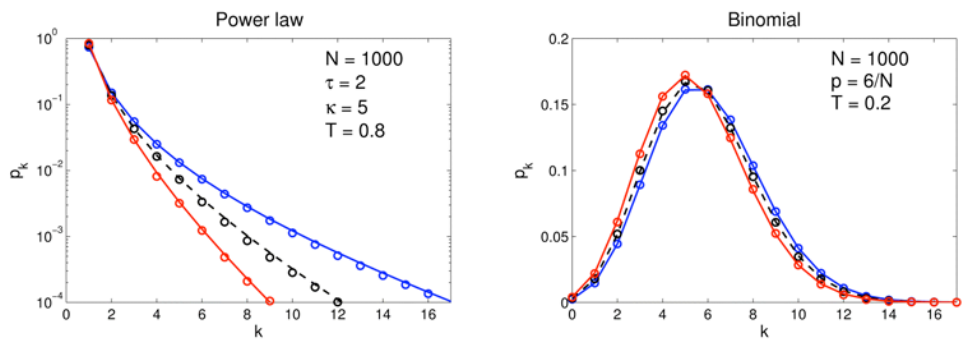

**Figure 3: Time evolution of typical degree distributions representing susceptible individuals.** The evolution of the degree distribution of the susceptibles is shown for 2 typical networks: Left) the power-law distribution used in figure 2; Right) a binomial distribution with $\boldsymbol{p} = 6/(N\text{-}1)$ and $\boldsymbol{N} = 1000$. The curves correspond to outbreak/epidemic sizes, $\boldsymbol{s}$, of 0 (solid blue), 200 (dashed black) and 400 (solid red). The transmissibility values were solely used to produce numerical results (circles).

### B. Additional failure of transmission

For the diseases considered in this paper (SEIR process), an individual can only be infected once; during their infectious period, an individual has the chance to transmit the disease to his/her neighbours, and is then permanently removed from the dynamics. This implies that as disease spreads, an infectious node is less likely to find susceptible nodes as they are already infected.

In the infinite-size limit, when the infectious vertices have a total number of $\hat{a}$ excess degrees, one expects the probability that these lead to $\hat{l}$ new infections will be

$$P(\hat{l}\,|\,\hat{a}) = \binom{\hat{a}}{\hat{l}} T^{\hat{l}} (1-T)^{\hat{a}-\hat{l}}$$. However, in the finite-size network, three main effects can

lead to the failure of potential transmissions: as shown in Figure 4, some of the stubs can lead to other infectious people, recovered individuals, or susceptible individuals who have already been targeted by another infectious stub.

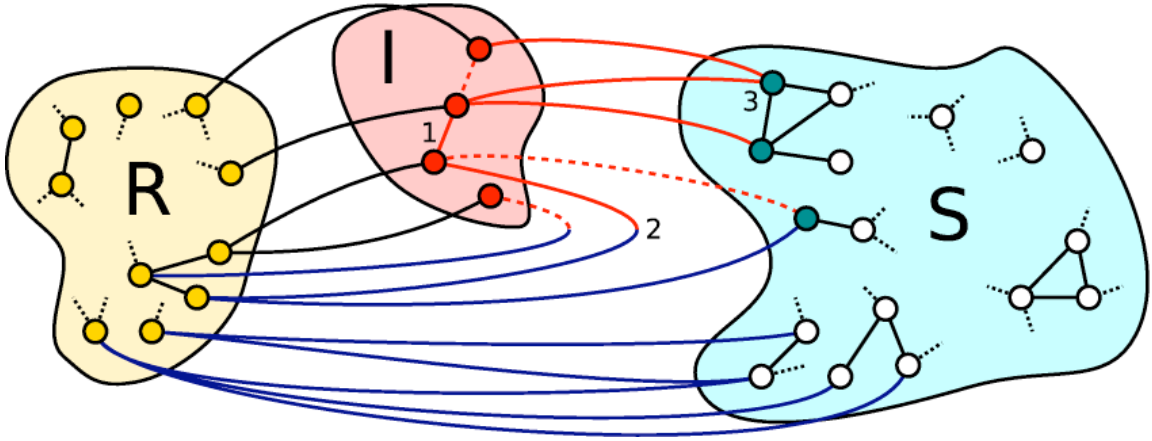

**Figure 4: Patterns leading to additional failure of transmission.** Infectious vertices (red circles) can transmit the disease along their stubs, other than the one they acquired the disease from; there are a total of $\hat{a}$ such stubs (excess degrees, red solid and dashed lines). In the infinite limit, each of these stubs have a probability, $T$, of leading to a new infection and a total of $\hat{l}$ will actually do so (red solid lines). Because past infectious nodes (now removed, yellow circles) may have failed to transmit the disease to each of their susceptible neighbours, links between recovered (yellow circles) and susceptibles (cyan and white circles) or infectious are not forbidden (blue solid lines, total of $\hat{b}$). In finite cases, some of the $\hat{l}$ stubs can: 1) target another infectious vertex; or 2) be part of the $\hat{b}$ edges that are not forbidden to link infectious to recovered; or 3) target a susceptible already targeted by another infectious. Each of these possibilities leads to additional transmission failure that must be considered when the network is finite.

When in addition to $\hat{a}$ and $\hat{l}$, one also knows the number, $\hat{n}$, of stubs belonging to the susceptibles and number, $\hat{b}$, of links joining the recovered to the susceptible or infectious nodes (excluding the links that transmitted the disease to them; see Figure 4), a differential equation can be solved in the mean-field approximation. This equation enables one to obtain the distribution of *m* new infections

$$P(m\,|\,\hat{a},\hat{b},\hat{l},\hat{n}) = \binom{\hat{l}}{m} \rho^{\hat{l}} (1-\rho)^{m-\hat{l}}$$, where

$$\rho = \frac{(N-s')}{\widehat{l}} \left[ 1 - G_0^s \left( 1 - \frac{\widehat{l}}{\widehat{a} + \widehat{b} + \widehat{n}} ; s' \right) \right]. \tag{16}$$

Therefore, the new forward recurrence relation, which takes into account both the evolution of the degree distribution and the additional failure of transmission, can be written as

$$\Psi_0^g(x,y) = \sum_{s',m'} \psi_{s',m'}^{g-1} \cdot x^{s'} \left[ G_1^l \left( 1 - (xy-1)T\rho(s',m'); s',m' \right) \right]^{m'}. \tag{17}$$

For more details on the derivation of these equations, please see Supplementary Online Material. Figure 5 depicts the phase-space representation for the finite-size power-law network, using the abovementioned equation along with its projection on the **m**-axis.

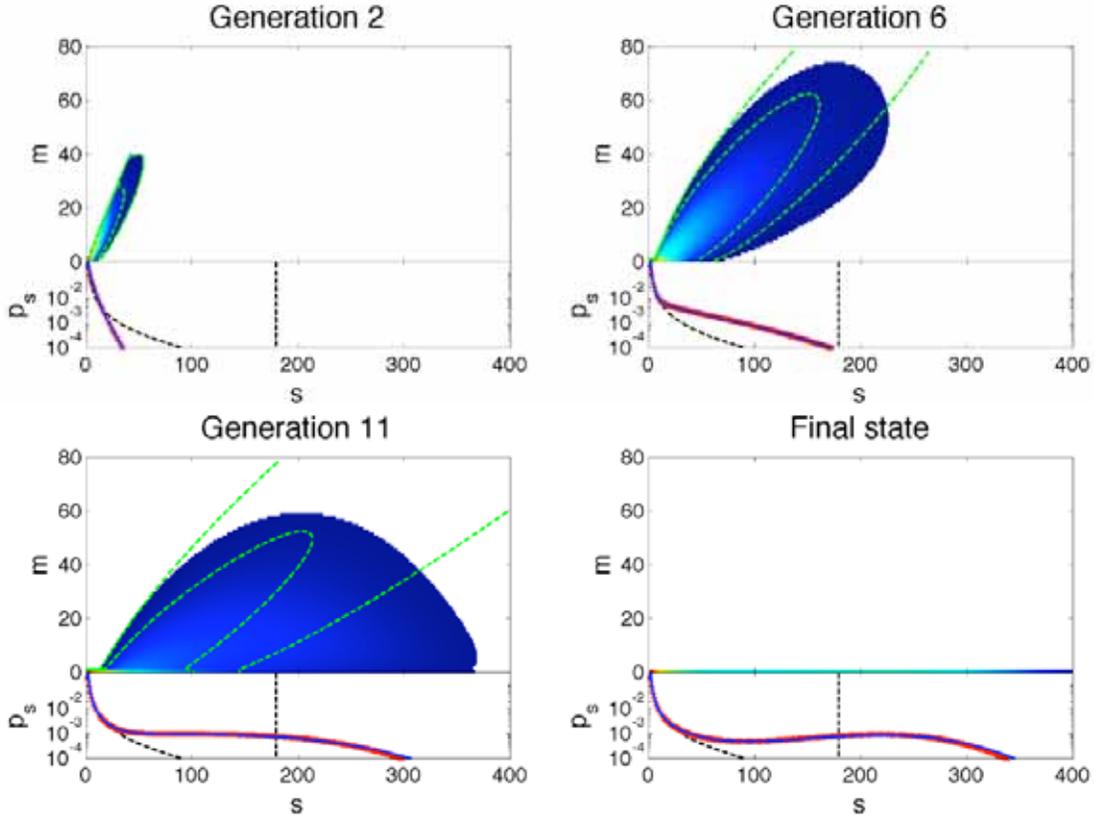

**Figure 5: Phase-space representation for the finite-size network algorithm.** The same situation as in figure 2 is displayed, but this time we used equation (17) instead of (11). The green dashed curves in the phase-space diagrams are contour plots of the previous results for infinite-size estimates (figure 2). The results produced by the finite-size algorithm are in very good agreement with the numerical simulations (red crosses) over the entire range of possible outbreak/epidemic sizes. The numerical results are identical to those shown in figure 2. While the time-independent formalism produces a single number for the giant component size (represented by the vertical dashed line), the time-dependent formalism produces the whole probability distribution of sizes above epidemic threshold.

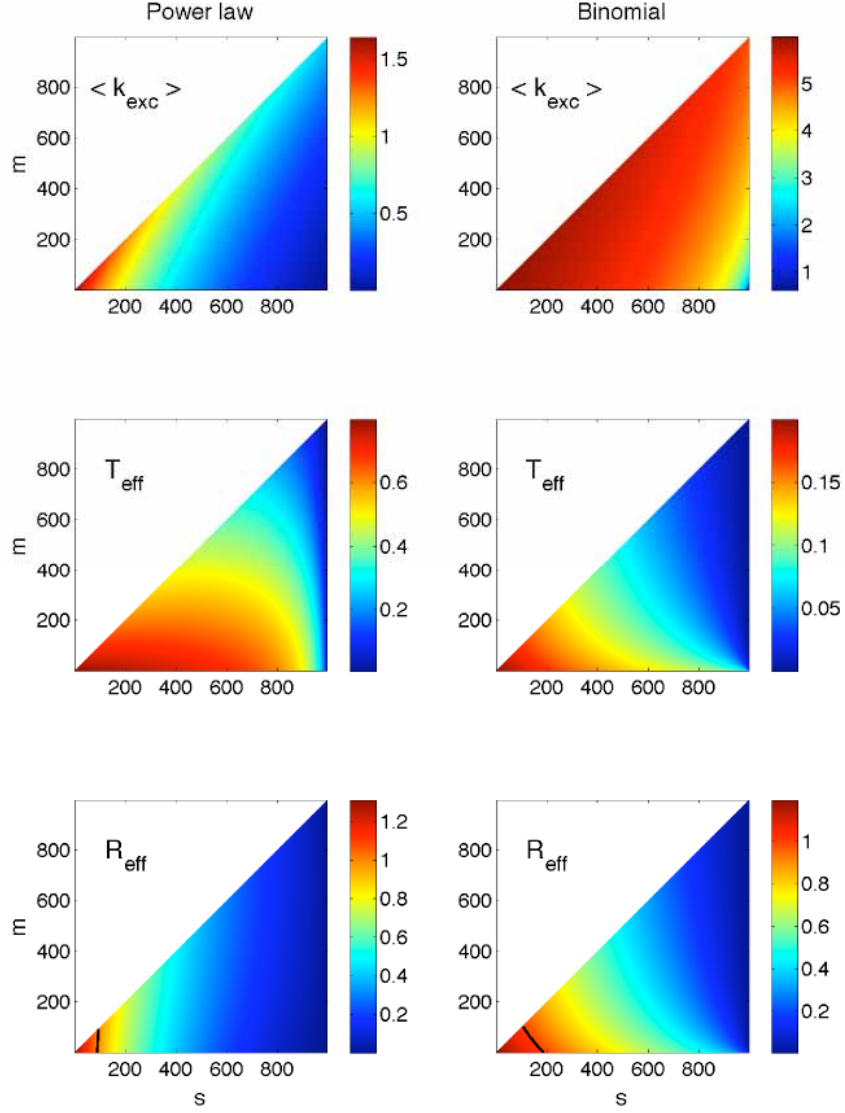

**Figure 6: "Effective reproduction number" interpretation.** For the two networks presented in figure 3 we show the expected excess degree of the infectious $\langle k_{exc} \rangle$, the effective transmissibility $T_{eff}$ and the corresponding effective reproduction number $R_{eff}$ for each $(s, m)$ state. The solid black line in the $R_{eff}$ plots corresponds to the "threshold" value $R_{eff} = 1$.

In order to establish a link between this formalism and the classical epidemiological models, it is worth revisiting the interpretation of the basic reproductive number – a key parameter in the classical epidemiology[1, 16] – based on the formalism offered in this paper. Using equations (15) and (16) one can derive the expected excess degree of the infectious nodes, $\langle k_{exc} \rangle = G_1'^I(1; s, m)$, and the effective transmissibility $T_{eff} = T \rho(s, m)$.

One can then define the corresponding *effective reproductive number* $R_{eff} = \langle k_{exc} \rangle T_{eff}$ for each $(s, m)$ state. Figure 6 shows the dependency of $\langle k_{exc} \rangle$, $\boldsymbol{T_{eff}}$, and $\boldsymbol{R_{eff}}$ on $\boldsymbol{s}$ and $\boldsymbol{m}$ for the networks introduced in figure 3. It is worth noting that the behaviour of the power law distribution is dominated by the variability of $\langle k_{exc} \rangle$, as $\boldsymbol{T_{eff}}$ remains uniform along the $\boldsymbol{s}$ axis, particularly in the vicinity of the epidemic threshold. However, the converse is true for the binomial distribution; it is the variability of $\boldsymbol{T_{eff}}$ that is responsible for the behaviour of this distribution.

## Conclusion

The emergence and re-emergence of infectious diseases pose a great threat to public health. The potential spread of a new pandemic strain of influenza or other emerging infection, such as SARS, may have a devastating impact on human lives and economies. There is an urgent need to develop reliable quantitative tools that can be used to compare the impact of various intervention strategies in real time. These tools must be able to incorporate the detailed structure of contact networks responsible for disease spread, as well as compare various intervention outcomes during the time of crisis in a relatively short time span. In addition, these tools should be as equally applicable to large-scale networks as to finite-size networks, seeing that many interventions must be implemented not only globally, but locally (e.g., hospital settings, schools) as well.

In this paper, we introduced and validated a theoretical framework that will enable one to incorporate these two important aspects of disease outbreaks/epidemics, simultaneously. The spread of infectious disease within a finite-size network is a complex phenomenon,with dynamics that will not likely be defined in an exact analytical manner. Interestingly, despite using a simple representation of this complex interaction – namely choosing $\boldsymbol{s}$ and $\boldsymbol{m}$ as the main ingredients within our formalism and expecting the mean value for every other detailed quantities – we demonstrated that an accurate image of these dynamics can be portrayed using a generation-based approach. In using this approach a continuous-time element is provided by way of mapping generations through time with the use of a continuous transmissibility function, $\boldsymbol{T_g(t)}$. A more comprehensive analytical framework, in principle, can be established on a more natural continuous-time basis, specifically, $\boldsymbol{s}$ and its time derivative, $\dot{s}$. Although a continuous-time analysis of any dynamical system is of utmost desire, as far as practical considerations are concerned, the proposed framework can provide valuable insights into many realistic situations of public health importance. Such insight was not attainable before this point.

## Acknowledgements

BP would like to acknowledge the support of the Canadian Institutes of Health Research (grants no. MOP-81273 and PPR-79231), the Michael Smith Foundation for Health Research (Senior Scholar funds) and the British Columbia Ministry of Health (pandemic preparedness modeling project).